\pdfoutput=1		

\documentclass[a4paper,12pt]{article}  
\usepackage[vmargin={20mm,20mm},hmargin={20mm,20mm}]{geometry}
\usepackage{commath}
\usepackage{graphicx}
\usepackage{enumitem}
\usepackage{tikz-cd}
\usepackage{subcaption}
\usepackage[american]{babel}

\usepackage{amsmath}
\usepackage{amssymb}
\usepackage{amsthm}
\usepackage{bbm}
\usepackage{mathrsfs}

\DeclareMathOperator{\SD}{SD}

\DeclareMathOperator{\Var}{Var}
\DeclareMathOperator{\cond}{cond}

\DeclareMathOperator{\sd}{sd}

\DeclareMathOperator{\Bias}{Bias}
\DeclareMathOperator{\MSE}{MSE}

\DeclareMathOperator{\MCSE}{MCSE}
\DeclareMathOperator{\clus}{clus}
\DeclareMathOperator{\elpd}{elpd}
\DeclareMathOperator{\logo}{logo}
\DeclareMathOperator{\SE}{SE}
\DeclareMathOperator{\SMD}{SMD}
\DeclareMathOperator{\red}{red}

\DeclareMathOperator{\Cov}{Cov}

\newtheoremstyle{main} 		             	 		
  	{}	                                     		
  	{}	                                    		
  	{\itshape}			                     		
  	{}        	                             		
  	{\boldmath\bfseries}   	                         		
  	{.}            	                        		
  	{ }           	                         		
  	{\thmname{#1}\thmnumber{ #2}\thmnote{ (#3)}}	
\theoremstyle{main}

\newtheorem{theorem}{Theorem}[section]

\newtheoremstyle{nonit} 		             	 		
  	{}	                                     		
  	{}	                                    		
  	{}			                     		
  	{}        	                             		
	{\boldmath\bfseries}   	                         		
  	{.}            	                        		
  	{ }           	                         		
  	{\thmname{#1}\thmnumber{ #2}\thmnote{ (#3)}}	
\theoremstyle{nonit}

\newtheoremstyle{ex} 		             	 		
  	{}	                                     		
  	{}	                                    		
  	{}			                     		
  	{}        	                             		
  	{\bfseries\boldmath}   	                         		
  	{.}            	                        		
  	{ }           	                         		
  	{\thmname{#1}\thmnumber{ #2}\thmnote{ (#3)}}	
\theoremstyle{ex}

\usepackage[backend=bibtex, style=numeric]{biblatex}
\bibliography{bibliography}
\usepackage[bookmarksopen=true,bookmarksnumbered=true]{hyperref}
\hypersetup{
  colorlinks   = true, 
  urlcolor     = blue, 
  linkcolor    = blue, 
  citecolor    = blue 
}

\usepackage{tcolorbox}
\tcbuselibrary{theorems}
\newtcbtheorem
  []
  {summary}
  {Summary}
  {%
    colback=blue!5,
    colframe=blue!35!black,
    fonttitle=\bfseries,
  }
  {sum}

\usepackage{booktabs}

\begin{document}

\title{Bayesian Cluster-Adjusted Standardised Mean Differences: A Simulation Study}

\author{Yannis B\"ahni\footnote{\href{mailto:yannis.baehni@gess.ethz.ch}{yannis.baehni@gess.ethz.ch}, MINT Lernzentrum ETH Z\"urich, RZ E9, Clausiusstrasse 59, 8092 Z\"urich, Switzerland}
}

\maketitle

\begin{abstract}
Reporting standardised mean differences of continuous psychometric measures is omnipresent in psychology. In a randomised controlled trial with a multilevel structure, one can define a cluster-adjusted version of these quantities naturally taking into account heterogeneous residual variances and accounting for the multilevel structure. In this article, we carefully define the underlying location-scale distributional mixed-effects models and derive cluster-adjusted standardised mean differences using partial pooling. Estimating these quantities is best handled using Bayesian statistics, and we perform an extensive Monte Carlo simulation study to test the performance of these metrics under different simulation configurations. We conclude that in order to measure the a-priori heterogeneous treatment effect across clusters in practically relevant scenarios, it is enough to consider random effects only on the location model and not on the scale model. Even under artificial severe heteroscedasticity, this reduced model performs well both in terms of parameter bias and predictive accuracy. The resulting cluster-adjusted standardised mean difference is an average value across clusters, where the heterogeneity is summarised via its standard deviation.
\end{abstract}

\tableofcontents

\section*{Impact Statement}
Proper handling of heterogeneity in any psychometric measurement is an important aspect of empirical research. Robust and unbiased estimation of heterogeneous residual variances and incorporation of the multilevel structure assure reliable estimates and increase replicability while simultaneously capturing realistic patterns. Bayesian multilevel models offer the necessary flexibility, and with Cohen's $d$ being among the most widely used standardised mean differences, one should carefully analyse its properties and limitations under this model-based framework. Therefore, we add a powerful Bayesian tool to the "new statistics" initiated by Geoff Cumming in 2014, moving beyond simple null-hypothesis significance testing and ensuring the integrity of future psychological research by estimating not only an average treatment effect but also simultaneously quantifying its heterogeneity across clusters.

\section{Introduction}
\begin{figure}[h!tb]
    \centering
    \includegraphics[width=\linewidth]{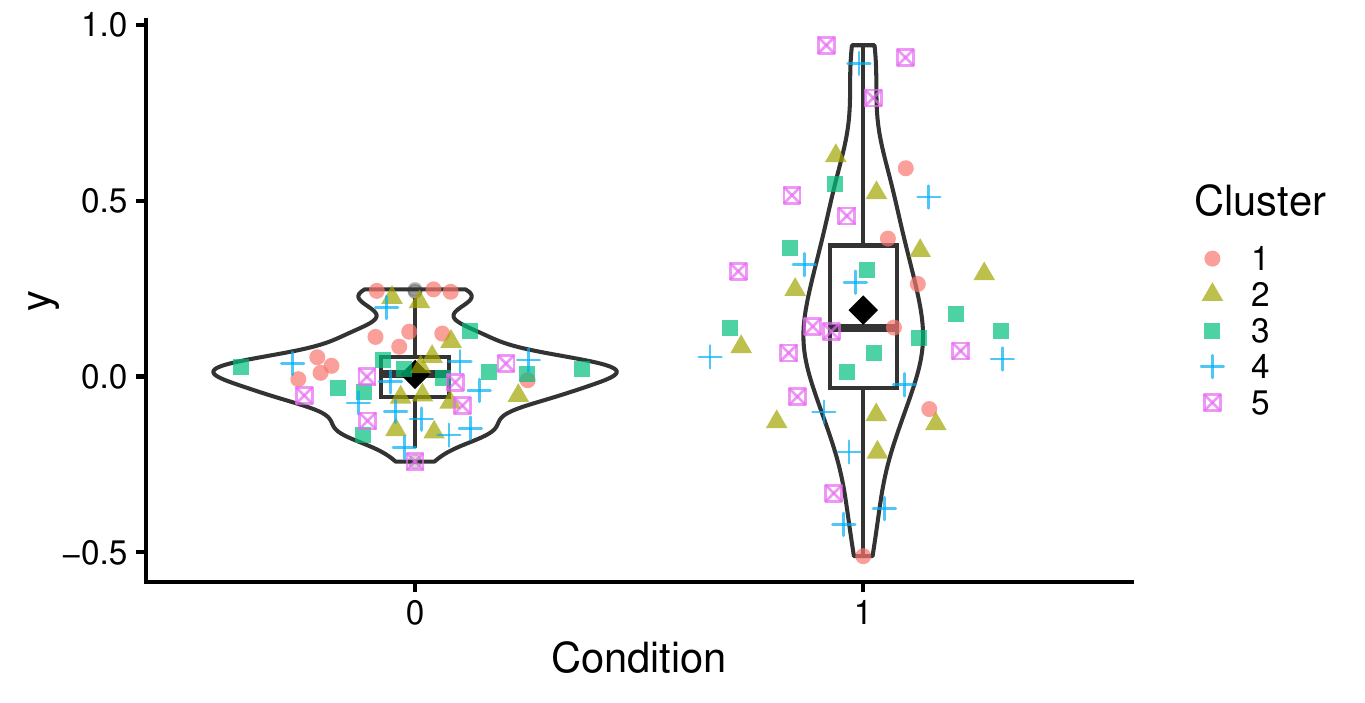}
    \caption{Distribution of the continuous response variable $y$ with respect to condition and clusters. Violin shapes indicate densities of $y$, overlaid with boxplots. Squared black points represent mean values. Shaped and coloured points represent the labelled members of the clusters}
    \label{fig:violin}
\end{figure}

Statistically quantifying the difference in group-means is ubiquitous in psychology. We consider a continuous response variable $y$ for two factorial conditions $i = 0, 1$ (control vs. treatment), as shown in Figure \ref{fig:violin}. Denote by $\mu_0$ and $\mu_1$ the group-means of the control and treatment conditions, respectively. These mean values represent the \emph{location} for the two conditions. The \emph{scale} of the two conditions is quantified by the standard deviations $\sigma_0$ and $\sigma_1$. Therefore, the corresponding location-scale model \cite{buerkner:advanced:2018, stasinopoulos:gam:2024} is given by 
\begin{equation*}
    \mu_i = \beta_0 + \beta_1 \cdot \delta_{i1} \qquad \text{and} \qquad \log(\sigma_i) = \gamma_i,
\end{equation*}
\noindent where we denote by
\begin{equation*}
    \delta_{kl} := \begin{cases}
        1 & k = l,\\
        0 & k \neq l
    \end{cases}
\end{equation*}
\noindent the Kronecker delta. A common standardised mean-difference representing the unit-free magnitude of the difference between the two conditions is provided by Cohen's
\begin{equation}
    \label{eq:d_s}
    d_s := \frac{\beta_1}{\sqrt{(1 - p)\sigma_0^2 + p\sigma_1^2}} = \frac{\beta_1}{\sqrt{(1 - p)e^{2\gamma_0} + pe^{2\gamma_1}}}, \qquad p := \frac{n_1 - 1}{n_0 + n_1 - 2}, 
\end{equation}
\noindent as in \cite{lakens:d:2013}. Here, $n_0$ and $n_1$ denote the number of observations in the control and treatment conditions, respectively, and $p$ is the probability of belonging to the treatment condition. Typically, observations are nested within clusters (for example, classes in a realistic school scenario). We assume that the condition is quasi-randomised within the clusters, as in Figure \ref{fig:violin}.

A statistical model for adequately quantifying the difference between the two group-means should incorporate both the location and scale of the data, as well as the multilevel structure. Therefore, a natural model to consider is a mixed effects location-scale model \cite{richardson:melsm:2025} or a double hierarchical generalised linear model \cite{lee:dhglm:2006, morales:dhglm:2022}. In such a model, not only are the group-means estimated with respect to the random structure via partial pooling, but also the a-priori heterogeneous variances. Therefore, the adjusted model is given by
\begin{align}
    \label{eq:location}\mu_{ij} &= \beta_0 + u_{0j}\cdot \delta_{i0} + (\beta_1 + u_{1j})\cdot \delta_{i1}\\
    \label{eq:scale}\log(\sigma_{ij}) &= (\gamma_0 + v_{0j})\cdot \delta_{i0} + (\gamma_1 + v_{1j}) \cdot \delta_{i1}.
\end{align}
The fixed effects $\beta_i$ and $\gamma_i$ are the grouped grand-means of the model, whereas the random effects $u_{ij}$ and $v_{ij}$ describe the normal deviations from these global estimates for each cluster $j$ in condition $i = 0,1$. The distributional location random effects satisfy
\begin{equation}
    \label{eq:sd_location}
    \begin{pmatrix}
        u_{0j}\\
        u_{1j}
    \end{pmatrix} \sim N\left(\begin{pmatrix}
        0\\
        0
    \end{pmatrix}, \begin{pmatrix}
        \sd_0^2 & \rho_u \sd_0\sd_1\\
        \rho_u \sd_0\sd_1 & \sd_1^2
    \end{pmatrix}\right)
\end{equation}
\noindent for between-cluster variances $\sd_i^2$ and the random-slope correlation $\rho_u$ for the two conditions. The distributional scale random effects satisfy
\begin{equation}
    \label{eq:sd_scale}
    \begin{pmatrix}
        v_{0j}\\
        v_{1j}
    \end{pmatrix} \sim N\left(\begin{pmatrix}
        0\\
        0
    \end{pmatrix}, \begin{pmatrix}
        \sd_{\sigma_0}^2 & \rho_v \sd_{\sigma_0}\sd_{\sigma_1}\\
        \rho_v \sd_{\sigma_0}\sd_{\sigma_1} & \sd_{\sigma_1}^2
    \end{pmatrix}\right),
\end{equation}
\noindent for between-cluster variances $\sd_{\sigma_i}^2$ and the random-slope correlation $\rho_v$ for the two conditions. For defining a cluster-adjusted standardised mean difference, we first compute the marginal variance of this model across the entire population.

\begin{theorem}
    \label{thm:marginal_variance}
    For the location-scale distributional mixed-effects model \eqref{eq:location}, \eqref{eq:scale}, \eqref{eq:sd_location}, and $\eqref{eq:sd_scale}$, the marginal variance of the continuous response variable $y$ is given by
\begin{equation*}
    \Var(y) = (1 - p)\left(e^{2\gamma_0 + 2\sd_{\sigma_0}^2} + \sd_0^2 \right) + p\left(e^{2\gamma_1 + 2\sd_{\sigma_1}^2} + \sd_1^2 \right) + p(1 - p)\beta_1^2,
\end{equation*}
\noindent for the probability $p \in (0, 1)$ of belonging to the treatment condition.
\end{theorem}

\begin{proof}
    By the law of total variance, we have to compute
    \begin{equation*}
        \Var(y) = \mathbb{E}[\Var(y\vert \cond, \clus)] + \Var(\mathbb{E}[y\vert \cond, \clus]).
    \end{equation*}
    For a fixed condition $i = 0,1$, we compute
    \allowdisplaybreaks
    \begin{align*}
        \mathbb{E}[\Var(y\vert \cond = i, \clus)] &= \mathbb{E}[\sigma_{ij}^2] & (\text{condition is fixed})\\
        &= \mathbb{E}[e^{2\gamma_i + 2v_{ij}}] & (\text{scale model \eqref{eq:scale}}) \\
        &= e^{2\gamma_i}\mathbb{E}[e^{2v_{ij}}] & (\text{linearity of expectation})\\
        &= e^{2\gamma_i}e^{2\sd_{\sigma_i}^2} & (2v_{ij} \sim N(0, 4\sd_{\sigma_i}^2) \text{ by \eqref{eq:sd_scale}})\\
        &= e^{2\gamma_i + 2\sd_{\sigma_i}^2},
    \end{align*}
    \noindent since $\mathbb{E}[e^X] = e^{\mu + \frac{1}{2}\sigma^2}$ for $X \sim N(\mu, \sigma^2)$. For the other term, we compute
    \begin{equation*}
        \Var(\mathbb{E}[y\vert \cond = i, \clus]) = \Var(\mu_{ij}) = \Var(\beta_0 + \beta_1\cdot \delta_{i1} + u_{ij}) = \Var(u_{ij}) = \sd_i^2
    \end{equation*}
    \noindent using \eqref{eq:location}, since $u_{ij} \sim N(0, \sd_i^2)$ by \eqref{eq:sd_location}. Therefore, we get the group-specific marginal variances
    \begin{equation*}
        V_i := \mathbb{E}[\Var(y\vert \cond = i, \clus)] +  \Var(\mathbb{E}[y\vert \cond = i, \clus]) = e^{2\gamma_i + 2\sd_{\sigma_i}^2} + \sd_i^2
    \end{equation*}
    \noindent for $i = 0,1$. As $\cond$ is binary, we finally deduce
    \begin{equation*}
        \mathbb{E}[\Var(y\vert \cond, \clus)] = (1 - p)\cdot V_0 + p\cdot V_1 \quad \text{and} \quad \Var(\mathbb{E}[y\vert \cond, \clus]) = p\cdot(1 - p)\cdot\beta_1^2,
    \end{equation*}
    \noindent since $\mathbb{E}[y\vert \cond, \clus]$ is a binary discrete random variable such that $\mathbb{E}[y\vert \cond = 0, \clus] = \beta_0$ with probability $1 - p$ and $\mathbb{E}[y\vert \cond = 1, \clus] = \beta_0 + \beta_1$ with probability $p$.
\end{proof}

Using the marginal variance computed in Theorem \ref{thm:marginal_variance}, we can define a standardised mean difference for the model \eqref{eq:location}, \eqref{eq:scale}, \eqref{eq:sd_location}, and $\eqref{eq:sd_scale}$ by
\begin{equation}
    \label{eq:SMD}
    \SMD := \frac{\beta_1}{\sqrt{\Var(y) - p(1 - p)\beta_1^2}}.
\end{equation}
Removing the between-condition variance $p(1 - p)\beta_1^2$  allows us to compare the two conditions independently. Indeed, otherwise, a large difference $\beta_1$ between the two conditions would be penalised by a small standardised mean difference $\SMD$. The quantity $\SMD$ can now be defined for each cluster because we can simply use the numerator $\beta_1 + u_{1j} - u_{0j}$ for all $j$. However, for the standardiser in the denominator, there are multiple options for including the scale estimates \eqref{eq:scale} and \eqref{eq:sd_scale} with different interpretations. We note that 
\begin{equation*}
    \lim_{\sd_i, \sd_{\sigma_i} \to 0}\SMD = \frac{\beta_1}{\sqrt{(1 - p)e^{2\gamma_0} + pe^{2\gamma_1}}}, \qquad p := \frac{n_1 - 1}{n_0 + n_1 - 2}, 
\end{equation*}
\noindent is precisely Cohen's $d_s$ as in \eqref{eq:d_s} in the case of complete pooling.

\subsection{The Purely Within-Cluster Approach}
In this approach, one defines
\begin{equation}
    \label{eq:SMD_W}
    \SMD_{W}(j) := \frac{\beta_1 + u_{1j} - u_{0j}}{\sqrt{(1 - p)e^{2\gamma_0 + 2v_{0j}} + pe^{2\gamma_1 + 2v_{1j}}}}
\end{equation}
\noindent for each cluster $j$. The quantity $\SMD_{W}(j)$ measures the treatment effect relative only to the internal noise of the specific cluster $j$. Therefore, this highly localised standardised mean difference is suitable if classes naturally operate on completely different scales or environments. However, if a cluster has near-zero internal variation, $\SMD_{W}(j)$ will be artificially inflated, even if the raw effect is small.

\subsection{The Marginal Variance Approach}
In this approach, one defines
\begin{equation}
    \label{eq:SMD_H}
    \SMD_V(j) := \frac{\beta_1 + u_{1j} - u_{0j}}{\sqrt{(1 - p)(e^{2\gamma_0 + 2\sd_{\sigma_0}^2} + \sd_0^2)  + p(e^{2\gamma_1 + 2\sd_{\sigma_1}^2} + \sd_1^2)}} = \frac{\beta_1 + u_{1j} - u_{0j}}{\sqrt{\Var(y) - p(1 - p)\beta_1^2}}
\end{equation}
\noindent for each class $j$. The quantity $\SMD_V(j)$ measures the local treatment effect scaled against a global benchmark representing the whole population. Differences for $\SMD_V(j)$ are driven purely by differences in the raw treatment effect because the denominator is identical for all classes. However, this stable and comparable quantity ignores whether a specific cluster was highly chaotic or highly controlled internally.

\subsection{The Maximal Approach}
In this approach, one defines
\begin{equation}
    \label{eq:SMD_M}
    \SMD_{M}(j) := \frac{\beta_1 + u_{1j} - u_{0j}}{\sqrt{(1 - p)(e^{2\gamma_0 + 2v_{0j} + 2\sd_{\sigma_0}^2} + \sd_0^2)  + p(e^{2\gamma_1 + 2v_{1j} + 2\sd_{\sigma_1}^2} + \sd_1^2)}}
\end{equation}
\noindent for each class $j$. The quantity $\SMD_{M}(j)$ measures the local treatment effect scaled against a benchmark that includes both global variance and local variance inflation. Therefore, this quantity accounts for both the baseline differences between clusters and the unpredictable fluctuations of noise. Thus, the maximal cluster-adjusted standardised mean difference $\SMD_M(j)$ is interpretatively complex as it blends two levels of variation. The standardised mean difference $\SMD_M(j)$ was used in \cite{baehni:d:2025} for analysing the heterogeneity of the treatment effect in a realistic scenario.

\subsection{The Mean and Standard Deviation Across Clusters}
The standardisers \eqref{eq:SMD_W}, \eqref{eq:SMD_H}, and \eqref{eq:SMD_M} have different statistical properties.

\begin{theorem}
    \label{thm:population}
    Consider the location-scale model \eqref{eq:location}, \eqref{eq:scale}, \eqref{eq:sd_location}, and \eqref{eq:sd_scale}. Then we compute
    \begin{equation*}
        \mathbb{E}[\SMD_V(j)] = \SMD \qquad \text{and} \qquad         \Var(\SMD_V(j)) = \frac{\sd_0^2 + \sd_1^2 - 2\rho_u \sd_0 \sd_1}{\Var(y) - p(1 - p)\beta_1^2}.
    \end{equation*}
     Moreover, if we assume that the random effects on the location $(u_{0j}, u_{1j})$ are independent of the random effects on the scale $(v_{0j}, v_{1j})$, then we have that $|\mathbb{E}[\SMD_W(j)]| \geq |\SMD|$ and
    \begin{equation*}
        |\mathbb{E}[\SMD_M(j)]| \geq \frac{|\beta_1|}{\sqrt{(1 - p)(e^{2\gamma_0 + 4\sd_{\sigma_0}^2} + \sd_0^2)  + p(e^{2\gamma_1 + 4\sd_{\sigma_1}^2} + \sd_1^2)}}.
    \end{equation*}
\end{theorem}

\begin{proof}
    We first show the estimate for $\mathbb{E}[\SMD_M(j)]$. We abbreviate
    \begin{equation*}
        \SMD_M(j) = \frac{\Delta_j}{\sqrt{V_j}}, \qquad \Delta_j := \beta_1 + u_{1j} - u_{0j},
    \end{equation*}
    \noindent and
    \begin{equation*}
        V_j := (1 - p)\left(e^{2\gamma_0 + 2v_{0j} + 2\sd_{\sigma_0}^2} + \sd_0^2\right)  + p\left(e^{2\gamma_1 + 2v_{1j} + 2\sd_{\sigma_1}^2} + \sd_1^2\right).
    \end{equation*}
    Since $(u_{0j}, u_{1j})$ and $(v_{0j}, v_{1j})$ are independent by assumption, we estimate
    \begin{equation*}
        |\mathbb{E}[\SMD_M(j)]| = \left|\mathbb{E}[\Delta_j] \cdot \mathbb{E}\left[ \frac{1}{\sqrt{V_j}} \right]\right| = |\beta_1| \cdot \mathbb{E}\left[ \frac{1}{\sqrt{V_j}} \right] \geq \frac{|\beta_1|}{\sqrt{\mathbb{E}[V_j]}}
    \end{equation*}
    \noindent by Jensen's inequality. This yields the desired inequality by computing $\mathbb{E}[V_j]$ similar to the proof of Theorem \ref{thm:marginal_variance}. By slightly adjusting this argument, we also get the the estimate $|\mathbb{E}[\SMD_W(j)]| \geq |\SMD|$ and that $\mathbb{E}[\SMD_V(j)] = \SMD$ since the denominator $\sqrt{V}$ is constant in that case. Finally, we compute 
    \allowdisplaybreaks
    \begin{align*}
        \Var(\SMD_V(j)) &= \frac{1}{V}\Var(\Delta_j)\\ 
        &= \frac{1}{V}\Var(u_{1j} - u_{0j})\\ 
        &= \frac{1}{V}\left(\Var(u_{1j}) + \Var(u_{0j}) - 2\Cov(u_{1j}, u_{0j})\right)\\
        &= \frac{1}{V}\left(\sd_1^2 + \sd_0^2 - 2\rho_u \sd_0^2\sd_1^2\right).
    \end{align*}
    This shows the stated formula for the variance of $\SMD_V(j)$.
\end{proof}

In practice, the population values in Theorem \ref{thm:population} are approximated by the finite-sample quantities
\begin{equation*}
    \SMD_* := \frac{1}{J}\sum_{j = 1}^J \SMD_*(j) \qquad \text{and} \qquad \SD_* := \sqrt{\frac{1}{J - 1} \sum_{j = 1}^J(\SMD_*(j) - \SMD_*)^2},
\end{equation*}
\noindent being the mean value and standard deviation across clusters $j = 1,\dots, J$, respectively. These aggregated meta-summary metrics are computed for each individual posterior draw and are summarised by their posterior mean, together with the corresponding credible interval \cite{Kruschke:ttest:2012, baehni:d:2025}. Moreover, these metrics have different properties and interpretations concerning the three standardisers:
\begin{itemize}
    \item For the purely within-cluster approach via $\SMD_W(j)$, the mean $\SMD_W$ is an average of local signal-to-noise ratios that can be heavily distorted by outlier clusters with tiny variances. The standard deviation $\SD_W$ reflects two confounding measures: how much the raw treatment effect varies and how much local noise varies.
    \item For the marginal variance approach via $\SMD_V(j)$, the mean $\SMD_V$ is the average treatment effect scaled by a constant population measure, and it is proportional to the fixed effect $\beta_1$. The standard deviation $\SD_V$ captures pure treatment effect heterogeneity and reflects how much the treatment impact varies across classes.
    \item For the maximal approach via $\SMD_M(j)$, the mean $\SMD_M$ is a conservative estimate for an overall population effect because it penalises the overall average for clusters with extreme noise or high model uncertainty. The standard deviation $\SD_M$ measures complex heterogeneity, as it blends treatment variation with local noise fluctuations, weighted by global variance constraints.
\end{itemize}
In summary, if we want to quantify the heterogeneity of the isolated treatment effect across clusters, we use the standard deviation $\SD_V$ of the average standardised mean difference $\SMD_V$. If we want to capture the realistic noise and variation of the environments in which the classes exist, we can consider the standard deviation $\SD_M$ of the average standardised mean difference $\SMD_M$. However, only in the marginal variance approach does the mean $\SMD_V = \frac{1}{J}\sum_{j = 1}^J \SMD_V(j)$ converge to the true population value $\SMD$ defined in \eqref{eq:SMD} as $J \to +\infty$. We note that both $\SMD_W$ and $\SMD_M$ also coincide with Cohen's $d_s$ in the case of complete pooling because the normal distribution $N(\mu, \sigma^2)$ becomes the Dirac delta function at zero for $\sigma \to 0$.

\subsection{A Word on Missing Data}
Missing data \cite{graham:missing:2012} is unavoidable in a realistic setting, where non-zero attrition rates are encountered either naturally or artificially. Bayesian models offer three ways to handle missing data under the missing at random assumption: multiple imputation of missing data \emph{before} model fitting, imputation of missing data \emph{during} model fitting, or imputation of missing response variables \emph{after} model fitting using the posterior predictions of the complete responses. Missing data in the location-scale model \eqref{eq:location}, \eqref{eq:scale}, \eqref{eq:sd_location}, and \eqref{eq:sd_scale} can be imputed after the model fitting of the complete responses because all predictor variables are complete \cite{campbell:missing:2026}. This is the simplest mechanism for handling missing data imputation and leads to slightly lower statistical power. Therefore, attrition is not an interesting parameter to vary in this scenario.

\subsection{Research Questions}
A-priori, it might be the case that for defining the finite-sample quantity $\SMD_V$, it is unnecessary to use the full marginal variance of $y$. It could be sufficient to ignore the random slopes and intercepts on the scale model \eqref{eq:scale} and \eqref{eq:sd_scale}, because this model is rather complex and computationally demanding. In particular, estimating the random-slope correlation $\rho_v$ on the scale level is unstable. This leads to the possibly misspecified reduced standardised mean difference
\begin{equation*}
    \SMD_{V, \red}(j) := \frac{\beta_1 + u_{1j} - u_{0j}}{\sqrt{(1 - p)(e^{2\gamma_0} + \sd_0^2) + p(e^{2\gamma_1} + \sd_1^2)}}.
\end{equation*}
The bias of the mean $\SMD_{V, \red}$ across a finite number of clusters $j = 1,\dots, J$ compared to the true population value $\SMD$ should depend on the level of heterogeneity present in the data-generating mechanism. This is achieved by either varying parameters in the scale submodel \eqref{eq:scale} and \eqref{eq:sd_scale}, or by adjusting the probability $p$ of belonging to the treatment condition. This also has implications for $\SMD_W$ and $\SMD_M$. Indeed, if the reduced model adequately recovers $\SMD$, this parsimonious estimation would circumvent the need to consider the interpretationally complex quantities $\SMD_W$ and $\SMD_M$ entirely. 

\subsubsection{First Research Question}
How do the reduced standardised mean difference $\SMD_{V, \red}$ and $\SMD_V$ compare with the true cluster-adjusted population value \eqref{eq:SMD} of the location-scale model \eqref{eq:location}, \eqref{eq:scale}, \eqref{eq:sd_location}, and \eqref{eq:sd_scale}? We expect that the bias of the reduced estimate is positive compared to the heterogeneous true population value because the not fully adjusted pooled standard deviation might underestimate the actual variation in the data.  This is expected since, by adjusting Theorem \ref{thm:population}, we get $|\mathbb{E}[\SMD_{V,\red}(j)]| \geq |\SMD|$. In practice, this difference might be negligible and should depend on both the level of heterogeneity present and the prevalent probability of belonging to the treatment condition.

\subsubsection{Second Research Question}
Even if there is no systematic bias for both measures in Research Question 1, it might still be the case that the predictive accuracy of the two models differs when predicting an entirely new, unseen cluster $\widehat{j}$. The cluster-specific treatment effect is then given by
\begin{equation*}
    \beta_1 + u_{1} - u_0, \qquad (u_0, u_1) \sim N(0, \widehat{\Sigma}_u)
\end{equation*}
\noindent for the estimated between-cluster location covariance matrix $\widehat{\Sigma}_u$. The estimated between-cluster scale covariance matrix $\widehat{\Sigma}_v$ is used to sample $\log$-scale deviations $(v_0, v_1) \sim N(0, \widehat{\Sigma}_v)$. Therefore, the estimated cluster-adjusted standardised mean difference for the new cluster is given by
\begin{equation*}
    \SMD_W(\widehat{j}) = \frac{\beta_1 + u_{1} - u_0}{\sqrt{(1 - p)e^{2\gamma_0 + 2v_0} + pe^{2\gamma_1 + 2v_1}}}.
\end{equation*}
It might be sufficient to consider instead the reduced within-cluster standardised mean difference
\begin{equation*}
    \SMD_{W, \red}(\widehat{j}) = \frac{\beta_1 + u_{1} - u_0}{\sqrt{(1 - p)e^{2\gamma_0} + pe^{2\gamma_1}}}
\end{equation*}
\noindent by ignoring the random slopes and intercepts in the scale model \eqref{eq:scale} and \eqref{eq:sd_scale}. Consequently, we are interested in the predictive performance of the full location-scale model \eqref{eq:location}, \eqref{eq:scale}, \eqref{eq:sd_location}, and \eqref{eq:sd_scale} against the reduced model without random slopes and intercepts at the scale level.

\section{Method}
We follow the planning and reporting structure for simulation studies in methodological research in psychology established in \cite{siepe:mcss:2024}. In particular, this involves transparency regarding the aims, the data-generating mechanism, the estimands and other targets, the methods, as well as the performance measures. The major aim of this study is to analyse the performance of the average standardised mean differences $\SMD_*$ under different specifications, as well as to assess its predictive accuracy under systematic model reduction. Therefore, we assess the bias of the estimation when compared to the true value of $\SMD$ in \eqref{eq:SMD} and use leave-one-group-out cross-validation to assess the predictive performance. All data are publicly available on Zenodo under [\href{https://zenodo.org/records/23831421?token=eyJhbGciOiJIUzUxMiJ9.eyJpZCI6Ijk1ZWFjYWZhLTQ2MmQtNDQzOC1hYzY4LTVhZTkwMWY4NGJhOCIsImRhdGEiOnt9LCJyYW5kb20iOiJhNzZhNGQyOGE2MDcxZTg1NWM5NDFhMDczMjI5N2IxOCJ9.lLI1ZPhsCITHZkeYzfvqK3dm7TbhqWMjVlK-kEkOWMxhDvvdHgt_zQ1Fpz82eNa6Ahz1maw9q_dI1yy38XEoXQ}{LINK}] with DOI \href{http://dx.doi.org/10.5281/zenodo.23831420}{10.5281/zenodo.23831420}.

\subsection{Data Generation}
We manually generated location-scale data in R \cite{R} for the model \eqref{eq:location}, \eqref{eq:scale}, \eqref{eq:sd_location}, and \eqref{eq:sd_scale} by varying several focal parameters. Overall, we fixed $\beta_0 := 0.00$, $\beta_1 := 0.10$, $\gamma_0 := -2.00$, $\sd_0 := 0.04$, $\sd_1 := 0.05$, $\rho_u = 0.70$, $\sd_{\sigma_0} := 0.10$, and $\rho_v = 0.20$ based on \cite{baehni:d:2025}. The other values now depended on different levels of heteroscedasticity for the treatment condition: None, moderate, and severe. The corresponding values are summarised in Table \ref{tab:hetero}. The probability $p$ of belonging to the treatment condition was varied to be $0.25$, $0.50$, or $0.75$, sampled from a binomial distribution. In the specifications for Research Question 1, the number of clusters was $15$, $30$, $45$, or $60$, where we sampled the cluster sizes from a Poisson distribution with $\lambda := 20$ and a minimal cluster size of $10$. This results in a $4 \times 3 \times 3 = 36$ fully factorial simulation design. For Research Question 2, we considered only $15$ and $60$ clusters under severe heterogeneity for a computationally intensive Bayesian predictive stress test. Indeed, if the full and reduced models do not differ in terms of predictive accuracy in this highly artificial setup, where the complex model should outperform the reduced one, this is also not true for more well-behaved settings.

\begin{table}[h!tb]
\caption{Specification of Parameters for the Three Different Heteroscedasticity Levels}
\begin{tabular}{lrrr}\midrule
Parameter & None & Moderate &
Severe\\\midrule
$\gamma_1$ & $-2.00$ & $-1.00$ & $-0.50$\\
$\sd_{\sigma_1}$ & $0.10$ & $0.20$ & $0.40$\\\midrule
\end{tabular}
\label{tab:hetero}
\end{table}

\subsection{Computational Aspects}
For Bayesian model estimation of the location-scale model, we used brms \cite{brms} with the fast backend cmdstanr \cite{cmdstanr}. The target distribution used was a default Gaussian distribution. Initial Bayesian diagnostic plots, such as (grouped) posterior density, (grouped) average residual error, (grouped) PIT-ECDF, and LOO-PIT-ECDF, showed that this adequately modelled the simulated data. The total number of iterations was $4000$, with $2000$ serving as a warmup on three parallel sampled Markov chains. This ensured adequate bulk and tail effective sample sizes (ESS) for all estimates of interest. The target acceptance rate was specified to be $0.99$, with a maximum tree depth of $15$. This resulted in R-hat estimates of $1.00$, with a negligible amount of divergent transitions below $0.05\%$. We used default improper flat prior distributions  for location and scale fixed effects, as well as half-Student-$t$ distributions with three degrees of freedom, location $0$ and scale $2.5$ for random effect standard deviations. Moreover, default LKJ priors with a shape parameter $1$ are used on the correlation matrices \cite{culpepper:LKJ:2026}. These default prior distributions for the group-level effects are weakly informative and improve Monte Carlo sampling \cite{merkle:bayes:2026, luedecke:bayes:2026}. Regarding hardware configurations, we used an Intel Core i9-10900K 3.7 GHz processor with 20 logical cores and 64 GB of RAM. Even with parallel computation via doFuture \cite{doFuture} and doRNG \cite{doRNG}, sampling required a considerable amount of time. To track the progress of the simulation, we used progressr \cite{progressr}. The main computational burden is that the leave-one-group-out cross-validation currently relies on exact refitting. This means that for each excluded cluster, the model is completely resampled. Therefore, we used three cores for parallelising the outer foreach-loop, and six inner worker cores for the $K$-fold grouped LOO. This heavy parallelisation is currently required, but future improvements might be on the way \cite{vehtari:logocv:2026}. Even with this optimised parallelisation, the runtime for each of the datasets with $60$ classes was four days for sampling $R = 480$ replications. Additionally, we used the R-packages ggbeeswarm \cite{ggbeeswarm} and ggplot2 \cite{ggplot2} for visualisation, as well as tidybayes \cite{tidybayes} and tidyverse \cite{tidyverse} for data management.

\subsection{Measures}
\subsubsection{Bias}
Given a fixed parameter $\theta$ and estimates $\widehat{\theta}_r$ for $r = 1, \dots, R$, one defines the parameter bias as the mean value of the differences among all replications via
\begin{equation*}
    \Bias(\widehat{\theta}, \theta) := \mu(\widehat{\theta}) - \theta = \frac{1}{R}\sum_{r = 1}^R (\widehat{\theta}_r - \theta), \qquad \mu(\widehat{\theta}) := \frac{1}{R} \sum_{r = 1}^R \widehat{\theta}_r.
\end{equation*}
The parameter bias gives the difference between the mean of the parameter estimates across the replications of the Monte Carlo simulation study and the true parameter value $\theta$. For a unit-free description, one can convert the bias to the \emph{percentage parameter bias}
\begin{equation}
    \label{eq:bias}
    \%\!\Bias(\widehat{\theta}, \theta) := \frac{100}{\theta}\Bias(\widehat{\theta}, \theta) =  \frac{100}{R}\sum_{r = 1}^R \frac{\widehat{\theta}_r - \theta}{\theta}
\end{equation}
 if $\theta \neq 0$, and $\%\!\Bias(\widehat{\theta}, \theta) := 100 \cdot \mu(\widehat{\theta})$ if $\theta = 0$. A common threshold for acceptable percentage parameter bias is within $\pm 10\%$. To quantify the sampling variance, we use the \emph{Monte Carlo standard error of the percentage bias} defined by 
\begin{equation*}
    \%\!\MCSE(\widehat{\theta}, \theta) := \frac{100}{\sqrt{R - 1}}\frac{\sd(\widehat{\theta})}{\theta}, \qquad \sd(\widehat{\theta}) := \sqrt{\frac{1}{R}\sum_{r = 1}^R(\widehat{\theta}_r - \mu(\widehat{\theta}))^2}.
\end{equation*}
An acceptable Monte Carlo standard error is below $5\%$ following \cite{siepe:mcss:2024}, ideally around $1\%$.

\subsubsection{Mean Squared Error}
Like the mean of the parameter estimates across the replications in the parameter bias, one can also measure its variance. One defines the mean squared error of the parameter estimation by
    \begin{equation*}
        \MSE(\widehat{\theta}, \theta) := \sd(\widehat{\theta})^2 + \Bias(\widehat{\theta}, \theta)^2.
    \end{equation*}
    A short computation shows that, equivalently, we have
    \begin{equation*}
        \MSE(\widehat{\theta}, \theta) = \frac{1}{R}\sum_{r = 1}^R(\widehat{\theta}_r - \theta)^2.
    \end{equation*}
    Again, a unit-free quantity derived from the mean squared error is the \emph{percentage mean squared error}
    \begin{equation}
        \label{eq:MSE}
        \%\!\MSE(\widehat{\theta}, \theta) := \frac{100}{\theta^2}\MSE(\widehat{\theta}, \theta) = \frac{100}{R}\sum_{r = 1}^R \left(\frac{\widehat{\theta}_r - \theta}{\theta}\right)^2.
    \end{equation}
    if $\theta \neq 0$, and $\%\!\MSE(\widehat{\theta}, \theta) := 100 \cdot \mu(\widehat{\theta}^2)$ if $\theta = 0$. An acceptable percentage mean squared error is below $10\%$.

\subsubsection{Difference in Predictive Accuracy}
Following \cite{vehtari:loo:2017, wagenmakers:loo:2019, vehtari:loo:2025, vehtari:logocv:2026}, we define the leave-one-group-out ($\logo$) expected predictive $\log$-density by
\begin{equation*}
    \elpd_{\logo}(\mathcal{M}|\clus) := \sum_{j = 1}^J \log \int \mathbb{P}_{\mathcal{M}}(\clus_j | \theta)\mathbb{P}_{\mathcal{M}}(\theta |\clus_{-j})d\theta,
\end{equation*}
\noindent for the model $\mathcal{M}$ and parameters $\theta$, where $\clus_{-j}$ means removing the $j$-th cluster from the data. Bayesian cross-validation proceeds by comparing the predictive performance for two not necessarily nested models $\mathcal{M}_0$ and $\mathcal{M}_1$. We consider the difference in their expected predictive performance
\begin{equation*}
    \Delta \elpd_{\logo}(\mathcal{M}_0, \mathcal{M}_1|\clus) := \elpd_{\logo}(\mathcal{M}_1|\clus) - \elpd_{\logo}(\mathcal{M}_0|\clus).
\end{equation*}
\noindent Via a normal approximation, we obtain the standard error
\begin{equation*}
    \SE_{\logo}(\mathcal{M}_0, \mathcal{M}_1|\clus) := \sqrt{\frac{J}{J - 1} \sum_{j = 1}^J \left(\Delta\elpd_{\logo, j}(\mathcal{M}_0, \mathcal{M}_1|\clus) - \overline{\Delta\elpd_{\logo}(\mathcal{M}_0, \mathcal{M}_1|\clus)}\right)^2},
\end{equation*}
\noindent for $\Delta \elpd_{\logo}(\mathcal{M}_0, \mathcal{M}_1|\clus)$, where
\begin{equation*}
    \overline{\Delta\elpd_{\logo}(\mathcal{M}_0, \mathcal{M}_1|\clus)} := \frac{1}{J}\sum_{k = 1}^J \Delta\elpd_{\logo, k}(\mathcal{M}_0, \mathcal{M}_1|\clus)
\end{equation*}
\noindent denotes the mean value, and for $k = 1,\dots, J$ we define
\begin{equation*}
    \Delta\elpd_{\logo, k}(\mathcal{M}_0, \mathcal{M}_1|\clus) := \log\frac{\displaystyle\int \mathbb{P}_{\mathcal{M}_1}(\clus_k | \theta_1)\mathbb{P}_{\mathcal{M}_1}(\theta_1 |\clus_{-k})d\theta_1}{\displaystyle\int \mathbb{P}_{\mathcal{M}_0}(\clus_k | \theta_0)\mathbb{P}_{\mathcal{M}_0}(\theta_0 |\clus_{-k})d\theta_0}
\end{equation*}
\noindent to be the $k$-th difference in the $\logo$ predictive $\log$-density for the $k$-th cluster $\clus_k$. There are no established hard thresholds for when either the model $\mathcal{M}_0$ or $\mathcal{M}_1$ robustly outperforms the other in terms of predictive accuracy, but a generally accepted threshold is
\begin{equation}
    \label{eq:logo}
    |\Delta \elpd_{\logo}(\mathcal{M}_0, \mathcal{M}_1|\clus)| > 4 \qquad \text{and} \qquad \frac{|\Delta \elpd_{\logo}(\mathcal{M}_0, \mathcal{M}_1|\clus)|}{\SE_{\logo}(\mathcal{M}_0, \mathcal{M}_1|\clus)} > 2.
\end{equation}
This threshold will be used to estimate the percentage $\%\!\logo$ of replications out of the number $R$ of all replications in which the full model outperforms the reduced one in terms of predictive accuracy when all but one cluster are used as training data. As currently no fast default procedure for computing the leave-one-group out cross-validation exists, we use a grouped $K$-fold approximation with $K = 12$ different partitions. This means that every model is refitted $12$ times, irrespective of the number of clusters. This is a trade-off regarding computational feasibility and precision and might be improved in future releases of brms \cite{vehtari:logocv:2026}. However, choosing $K = 12$, which is less than the number of distinct clusters, induces a slightly pessimistic estimation bias, meaning that the predictive performance is trained on fewer clusters, and thus the relative difference for the two models is lower than for the full number of clusters. This implies that the estimated percentage of replications where the full model outperforms the reduced one is slightly conservative. 

\section{Results}
\subsection{Research Question 1}
To answer the question of whether the reduced standardised mean difference $\SMD_{V, \red}$ and $\SMD_V$ exhibit a systematic bias compared to the true cluster-adjusted population value \eqref{eq:SMD}, Tables \ref{tab:sim_SMDV} and \ref{tab:sim_SMDVred} present the results of the simulation study for $\%\!\Bias$, $\%\!\MSE$, and $\%\!\MCSE$ concerning all simulation conditions. All reported results use $R = 480$ replications for each of the simulation conditions. The estimates $\SMD_V$ and $\SMD_{V, \red}$ denote the average finite-sample quantities over all clusters, and the true value of the data-generating model is provided by \eqref{eq:SMD}. 

\begin{table}
    \caption{Simulation Results for $\SMD_{V}$}
\begin{tabular}{lclcrr}\midrule
Heteroscedasticity & $p$ & $\#$Clusters & $\%\!\Bias$ & $\%\!\MSE$ &
$\%\!\MCSE$\\\midrule
None & $0.25$ & $15$ & $-4.91$ & $4.72$ & $0.97$\\
& & $30$ & $-3.12$  & $2.10$  & $0.65$\\
& & $45$ & $-0.91$ & $1.51$ & $0.55$\\
& & $60$ & $-0.50$  & $1.12$ & $0.47$\\
& $0.50$ & $15$ & $-3.89$ & $3.66$ & $0.86$\\
& & $30$ & $-2.48$  & $1.83$  & $0.61$\\
& & $45$ & $-1.23$ & $1.11$ & $0.47$\\
& & $60$ & $-0.48$ & $0.94$ & $0.43$\\
& $0.75$ & $15$ & $-4.30$  & $4.52$  & $0.95$\\
& & $30$ & $-1.07$ & $2.03$ & $0.65$\\
& & $45$ & $-1.17$  & $1.59$ & $0.56$\\
& & $60$ & $-0.51$ & $1.09$ & $0.47$\\\midrule
Moderate & $0.25$ & $15$ & $-2.51$ & $21.70$  & $2.12$\\
& & $30$ & $-0.88$ & $10.40$  & $1.47$\\
& & $45$ & $-2.06$ & $6.9$  & $1.17$\\
& & $60$ & $+0.91$ & $5.46$ & $1.05$\\
& $0.50$ & $15$ & $-2.57$ & $11.80$  & $1.56$\\
& & $30$ & $-1.70$ &  $5.51$ & $1.07$\\
& & $45$ & $-1.32$ & $4.08$ & $0.90$\\
& & $60$ & $-0.40$  & $3.21$ & $0.80$ \\
& $0.75$ & $15$ & $-4.90$  & $9.78$ & $1.41$\\
& & $30$ & $+0.32$ & $5.08$  & $1.03$\\
& & $45$ & $-0.39$ & $3.34$ & $0.82$\\
& & $60$ & $-0.91$ & $2.78$ & $0.75$\\\midrule
Severe & $0.25$ & $15$ & $-4.04$ & $47.30$  & $3.14$\\
& & $30$ & $-2.57$ & $25.40$  & $2.30$\\
& & $45$ & $-5.50$ & $19.2$  & $1.95$\\
& & $60$ & $-3.57$ & $12.2$  & $1.55$\\
& $0.50$ & $15$ & $-5.72$ & $24.90$  & $2.27$\\
& & $30$ & $-0.72$ & $13.00$   & $1.65$\\
& & $45$ & $-1.68$  & $9.13$ & $1.35$\\
& & $60$ & $-1.04$ & $6.99$ & $1.18$\\
& $0.75$ & $15$ & $-1.60$ & $18.20$  & $1.95$\\
& & $30$ & $-1.13$ & $9.39$  & $1.40$ \\
& & $45$ & $-1.23$ & $6.53$ & $1.14$\\
& & $60$ & $-2.10$  & $5.12$ & $1.01$\\\midrule
\end{tabular}
\label{tab:sim_SMDV}
\end{table}

\begin{table}
    \caption{Simulation Results for $\SMD_{V, \red}$}
\begin{tabular}{lcllrr}\midrule
Heteroscedasticity & $p$ & $\#$Clusters & $\%\!\Bias$ & $\%\!\MSE$ &
$\%\!\MCSE$\\\midrule
None & $0.25$ & $15$ & $-3.26$ & $4.64$  & $0.97$\\
& & $30$ & $-2.53$ & $2.09$  & $0.65$\\
& & $45$ & $-0.53$ & $1.53$ & $0.55$\\
& & $60$ & $-0.21$ & $1.13$ & $0.48$\\
& $0.50$ & $15$ & $-2.14$  & $3.65$ & $0.87$\\
& & $30$ & $-1.93$ & $1.81$ & $0.61$\\
& & $45$ & $-0.94$ & $1.10$  & $0.47$\\
& & $60$ & $-0.22$ & $0.95$ & $0.44$\\
& $0.75$ & $15$ & $-2.43$ & $4.54$ & $0.97$\\
& & $30$ & $-0.42$ & $2.04$ & $0.65$\\
& & $45$ & $-0.79$ & $1.63$ & $0.57$\\
& & $60$ & $-0.22$ & $1.10$  & $0.47$\\\midrule
Moderate & $0.25$ & $15$ & $+0.53$ & $23.8$  & $2.23$\\
& & $30$ & $+0.79$ & $11.0$  & $1.51$\\
& & $45$ & $-1.08$ & $7.19$ & $1.20$\\
& & $60$ & $+1.39$ & $5.96$ & $1.09$\\
& $0.50$ & $15$ & $-0.14$ & $12.70$  & $1.63$\\
& & $30$ & $-1.17$ & $5.74$ & $1.09$\\
& & $45$ & $-0.75$ & $4.31$ & $0.93$\\
& & $60$ & $-0.17$ & $3.28$ & $0.81$\\
& $0.75$ & $15$ & $-2.80$ & $10.40$  & $1.47$\\
& & $30$ & $+0.96$ & $5.35$ & $1.06$\\
& & $45$ & $+0.25$ & $3.56$ & $0.84$\\
& & $60$ & $-0.50$  & $2.90$  & $0.76$\\\midrule
Severe & $0.25$ & $15$ & $+3.87$ & $69.70$  & $3.81$\\
& & $30$ & $+0.01$ & $38.20$  & $2.82$\\
& & $45$ & $-4.27$  & $26.90$  & $2.32$\\
& & $60$ & $-2.03$ & $17.00$  & $1.84$\\
& $0.50$ & $15$ & $-1.15$ & $38.70$  & $2.84$\\
& & $30$ & $+1.07$ & $18.00$  & $1.94$\\
& & $45$ & $+0.47$  & $12.40$  & $1.58$\\
& & $60$ & $+0.79$ & $9.40$  & $1.37$\\
& $0.75$ & $15$ & $+3.55$ & $26.80$   & $2.36$\\
& & $30$ & $+2.66$ & $14.70$  & $1.75$\\
& & $45$ & $+0.52$ & $9.29$ & $1.36$\\
& & $60$ & $-1.69$ & $7.38$ & $1.21$\\\midrule
\end{tabular}
\label{tab:sim_SMDVred}
\end{table}

\subsection{Research Question 2}
For assessing the predictive performance of the full and the reduced model, we performed grouped $K$-fold cross-validations and used the criteria \eqref{eq:logo} to determine the percentage $\%\!\logo$ at which the full model outperforms the reduced one. The results of this stress-test under favourable conditions are summarised in Table \ref{tab:logo}. 

\begin{table}[h!tb]
\caption{Simulation Results for $\%\!\logo$ Under Severe Heterogeneity}
\begin{tabular}{lclr}\midrule
Heteroscedasticity & $p$ &$\#\text{Clusters}$ & $\%\!\logo$ \\\midrule
Severe & $0.25$ & $15$ & $5.83$\\
& & $60$ & $29.38$\\
& $0.50$ & $15$ & $11.04$\\
& & $60$ & $64.38$\\
& $0.75$ & $15$ & $17.50$\\
& & $60$ & $82.92$\\\midrule
\end{tabular}
\label{tab:logo}
\end{table}

\section{Discussion}
We compared two cluster-adjusted average standardised mean differences $\SMD_V$ and $\SMD_{V, \red}$ under different specifications to the true parameter value $\SMD$. In the first research question, we considered the parameter bias, mean squared error, and the Monte Carlo standard error. In all cases, the Monte Carlo standard error was below the threshold of $5\%$, implying reliable simulation results. This should be considered in light of the high computational demand from estimating Bayesian multilevel models, as well as $K$-fold grouped cross-validation, without any Pareto smoothed importance sampling (PSIS). We were still able to run $R = 480$ replications for each simulation setting in a reasonable time. Especially for leave-one-group-out cross-validation, there is room for improvement in computational efficiency. One should also note that the $\%\!\MCSE$ of $\%\!\Bias$ was slightly higher for the reduced model over all simulation settings compared to the full model.

As expected, for $15$ classes, the noise in the data-generating process was large for moderate and severe heteroscedasticity, which can be deduced from the corresponding percentage mean squared error in Table \ref{tab:sim_SMDV} and Table \ref{tab:sim_SMDVred}. This random noise drastically reduced for $60$ clusters. Except for $p = 0.25$, the $\%\!\MSE$ was well below the threshold $10\%$. In all cases and for both models, the percentage parameter bias was within $\pm 10\%$, indicating no large systematic bias in parameter estimation when compared to the true population value. This indicates that the maximal approach, as well as the possibly misspecified reduced approach, both recover the average standardised mean difference $\SMD$ essentially without bias. In particular, this implies that for estimating a cluster-adjusted standardised mean difference that captures the heterogeneity of the treatment effect across clusters, we can use $\SMD_{V, \red}$, as it is not strictly necessary or beneficial to also include scale random effects \eqref{eq:sd_scale}. Even under artificially severe heteroscedasticity, this reduced standardised mean difference performs well and yields robust estimates for reasonable cluster numbers. Furthermore, it is worth noting that the $\%\!\Bias$ is strictly smaller for the full model compared to the reduced model. This is in line with the theoretical observation $\mathbb{E}[\SMD_{V, \red}(j)] \geq \SMD$ following Theorem \ref{thm:population}. Since the standardised mean difference $\SMD_V$ slightly underestimates the true population value, this yields a bias around zero for $\SMD_{V,\red}$.

This conclusion is further strengthened by the difference in predictive performance under severe heterogeneity presented in Table \ref{tab:logo}. Even under artificially high heterogeneity, where the full model should outperform the reduced one in predicting clusters, the advantage quantified by the criterion \eqref{eq:logo} is limited. Only for $p = 0.75$ and $60$ clusters does the percentage of replications where the difference in the expected $\log$-predictive density is sufficiently large show $\#\logo$ larger than $80\%$. Even if $\#\logo$ was calculated only using $K = 12$ folds for those $60$ clusters, we can say that the criterion \eqref{eq:logo} is the bare minimum for measuring a robust model difference. Indeed, it is often recommended to use the stricter bound $|\Delta \elpd_{\logo}| > 4\SE_{\logo}$, for which $\#\logo$ drops down to $6.88\%$ under these more robust cross-validation criteria for $p = 0.75$ and $60$ clusters.

\subsection{Conclusions and Implications}
The results of this simulation study imply, or are at least compatible with, the conclusion that, in order to summarise and quantify the heterogeneity of a continuous psychometric measurement in a realistic scenario, it is sufficient to estimate the standardised mean difference $\SMD(j) := \SMD_{V,\red}(j)$. The underlying location-scale distributional mixed-effects model is given by 
\begin{equation*}
    \mu_{ij} = \beta_0 + \beta_1\cdot \delta_{i1} + u_{0j}\cdot \delta_{i0} + u_{1j}\cdot\delta_{i1}
\end{equation*}
\noindent with $\log(\sigma_{ij}) = \gamma_0\cdot \delta_{i0} + \gamma_1 \cdot \delta_{i1}$ and $(u_{0j}, u_{1j}) \sim N(0, \widehat{\Sigma}_u)$ for
\begin{equation*} 
 \widehat{\Sigma}_u = \begin{pmatrix}
        \sd_0^2 & \rho_u \sd_0\sd_1\\
        \rho_u \sd_0\sd_1 & \sd_1^2
    \end{pmatrix}.
\end{equation*}
This reduced model performs well in terms of predictive accuracy, measured by the percentage difference in $\elpd_{\logo}$ in Table \ref{tab:logo}. Even if this number is only an approximation of the true predictive accuracy, the results imply and support the conclusion that predicting an entirely new cluster $\widehat{j}$ via the reduced purely within-cluster standardised mean difference
\begin{equation*}
    \SMD(\widehat{j}) := \frac{\beta_1 + u_{1} - u_0}{\sqrt{(1 - p)e^{2\gamma_0} + pe^{2\gamma_1}}}, \qquad (u_0, u_1) \sim N(0, \widehat{\Sigma}_u).
\end{equation*}
Finally, we note that the reduced model yields a variance-decomposition for the corresponding true population \eqref{eq:SMD} value of the form
\begin{equation*}
    \SMD = \frac{\beta_1}{\sqrt{(1 - p)\sigma^2_0 + p\sigma_1^2 + (1 - p)\sd^2_0 + p\sd_1^2}} = \frac{\beta_1}{\sqrt{\sigma_W^2 + \sigma_B^2}}
\end{equation*}
\noindent for the within-cluster and between-cluster variances
\begin{equation*}
    \sigma_W^2 := (1 - p)\sigma^2_0 + p\sigma_1^2 \qquad \text{and} \qquad \sigma_B^2 := (1 - p)\sd^2_0 + p\sd_1^2,
\end{equation*}
\noindent respectively, in accordance with \cite{hedges:delta_t:2007}. In particular, this implies $|\SMD| < |d_s|$ for Cohen's $d_s$ as defined in \eqref{eq:d_s}, when both quantities are estimated from the proposed reduced model.

\subsection{Limitations and Future Research}
The first and obvious limitation is the approximated predictive accuracy via $K$-fold grouped cross-validation. Currently, this method relies on the exact refitting of all models, yielding a huge computational demand for large cluster numbers. Newly suggested approaches should be thoroughly tested first, and we encourage the computational improvement of $K$-fold grouped cross-validation. This could be studied well by systematically altering the simulation conditions in our study and including more specifications.

Another limitation is the study of different misspecifications of the data-generating mechanism. This means that one should include other distributions besides just Gaussian ones, as well as incorporate floor and ceiling effects. However, the overall limitation of the current approach is that it is still a variable-centred approach, meaning that no person-centred phenomena are captured, unlike in mixture modelling, for example. Additionally, the usual shortcomings of partial pooling also apply here.

An implication for future research is the computation of prediction intervals for entirely new, unseen clusters. These should be thoroughly assessed and validated. Prediction intervals quantify the inherent uncertainty in predicting the standardised mean difference in future studies. This should also be investigated in terms of the influence of prior distributions. This should be done for the suggested reduced location-scale model without random slopes and intercepts at the scale level. In that case, it is also interesting to study the finite-sample approximation of the standard deviation across clusters and compare it to the true population variance because estimating this variance directly might be unstable due to a large error associated with the random-slope correlation $\rho_u$ on the location level.

\section*{Code Availability}
All data are publicly available on Zenodo under [\href{https://zenodo.org/records/23831421?token=eyJhbGciOiJIUzUxMiJ9.eyJpZCI6Ijk1ZWFjYWZhLTQ2MmQtNDQzOC1hYzY4LTVhZTkwMWY4NGJhOCIsImRhdGEiOnt9LCJyYW5kb20iOiJhNzZhNGQyOGE2MDcxZTg1NWM5NDFhMDczMjI5N2IxOCJ9.lLI1ZPhsCITHZkeYzfvqK3dm7TbhqWMjVlK-kEkOWMxhDvvdHgt_zQ1Fpz82eNa6Ahz1maw9q_dI1yy38XEoXQ}{LINK}].
\addcontentsline{toc}{section}{References}
\printbibliography

\end{document}